\documentclass[a4paper]{article}
\RequirePackage[english]{babel}
\RequirePackage[latin1]{inputenc}
\RequirePackage[T1]{fontenc}
\RequirePackage{mathrsfs}
\RequirePackage{amsmath}
\RequirePackage{amssymb}
\RequirePackage{amsbsy}
\RequirePackage{bm}
\begin{document}
\title{\huge{\bf{Causality for ELKOs}}}
\author{Luca Fabbri\\
\footnotesize (INFN \& Theory Group, Department of Physics, University of Bologna, Italy)}
\date{}
\maketitle
\begin{abstract}
The importance for cosmology of the recently introduced ELKOs requires our deepest understanding of them and of all of their fundamental properties. Among these fundamental properties, a special one is causality: in the present paper, we show that causality is always preserved for ELKOs.
\end{abstract}
\section*{Introduction}
The standard model of cosmology has yet to settle many important issue, among which we find the following three: the first problem regards what could have caused the inflationary expansion that is supposed to have taken place soon after the big bang; the second problem concerns what could describe the dark sector of the matter distribution of the universe; the third problem demands the explanation for the axis of evil, a preferred axis that endows with a privileged direction the dynamics of the universe. Moreover, although these problems seem to be very different from one another, they do not only have to be addressed, but in addition they also have to be addressed within a unique scheme.

However hard this task might be, the aforementioned problems are likely to be solved altogether by a single theory recently proposed by Ahluwalia and Grumiller, defining a new form of matter called ELKO, autoconjugated spin-$\frac{1}{2}$ fields with mass dimension $1$ whose dynamics is constructed on second-order derivative lagrangians (see \cite{a-g/1}, \cite{a-g/2}); then the idea of a privileged direction in the dynamics arising from a preferred axis would be embraced by the spin structure of these special spinors (see \cite{a-l-s/1}, \cite{a-l-s/2}, \cite{l-m}, \cite{f-e}), whereas the curves of rotation of galaxies and the inflationary expansion of the universe would be reproduced by the richer dynamics of their lagrangian (see \cite{a}, \cite{s-s}, \cite{a-s}, \cite{b-s-i}, \cite{D}, \cite{b-m}, \cite{b/2}, \cite{b/11}, \cite{s}). Nevertheless, despite all the success it might enjoy, the quest for possible flaws of this theory is still open.

Among all the possible threats the theory of ELKOs could face is causality (see \cite{v-z}); if on the one hand it has recently been shown that ELKOs display causal behaviour under conditions that always exist (see \cite{f}), on the other hand knowing what these conditions exactly are would give the clearest insight about when such conditions hold, but more importantly when such conditions fail.

So in the present paper, we will explicitly find all the conditions of causality for ELKOs, showing first of all that it is always possible for them to hold, so that the result of \cite{f} is proven in the most general way, and by converse this will incidentally allow us to show that acausality can never be obtained, and therefore causality is always maintained for ELKOs.
\section{ELKOs}
The ELKOs are spinors with spin-$\frac{1}{2}$ transformation law that are eigenstates of the charge-conjugation operator $\lambda=\gamma^{2}\lambda^{*}$ up to a complex phase; their spin structure provides a mass dimension equal to $1$ and thus they must be described by second-order derivative lagrangians (all these results are derived in \cite{a-g/1} and \cite{a-g/2}, and they are reviewed in \cite{r-r}, \cite{r-h} and \cite{h-r}). Here we will consider the most general torsional case (the torsional generalization has been considered for instance in \cite{b-b/2}, \cite{b-b/1} and \cite{b/1}).

A generic ELKO $\lambda$ has its own definition of ELKO dual $\stackrel{\neg}{\lambda}$, as it is shown and justified in the aforementioned references; for them the derivatives $D_{\mu}$ are defined as usual, and in terms of the contorsionless derivatives $\nabla_{\mu}$ we have the decomposition 
\begin{eqnarray}
&D_{\mu}\lambda
=\nabla_{\mu}\lambda
+\frac{1}{2}K^{ij}_{\phantom{ij}\mu}\sigma_{ij}\lambda\\
&D_{\mu}\stackrel{\neg}{\lambda}
=\nabla_{\mu}\stackrel{\neg}{\lambda}
-\frac{1}{2}\stackrel{\neg}{\lambda}\sigma_{ij}K^{ij}_{\phantom{ij}\mu}
\label{derivatives}
\end{eqnarray}
in terms of the the $\sigma$ matrices and the contorsion tensor $K^{\rho}_{\phantom{\rho}\mu\nu}$, and the curvature tensor $G^{\rho}_{\phantom{\rho}\eta\mu\nu}$ is defined as usual: the contorsion tensor has one independent contraction given by $K_{\nu\rho}^{\phantom{\rho\nu}\rho}=K_{\nu}$ which we shall call Cartan vector, whereas the curvature tensor has only one independent contraction which is then given by $G^{\rho}_{\phantom{\rho}\eta\rho\nu}=G_{\eta\nu}$ which we shall call Ricci tensor, with only one contraction given by $G_{\eta\nu}g^{\eta\nu}=G$ called Ricci scalar. Finally, by taking the product of two derivatives of the spinor and the Ricci scalar one builds the general second-order derivative lagrangian, respectively for the ELKO and for the gravitational-torsional fields.

\subsection{ELKOs: features of differential equations}
The second-order derivative lagrangian is given by the simple
\begin{eqnarray}
L=G+D_{\mu}\stackrel{\neg}{\lambda}D^{\mu}\lambda-m^{2}\stackrel{\neg}{\lambda}\lambda
\label{lagrangian}
\end{eqnarray}
with $m$ mass of the field (for details about this lagrangian see \cite{a-g/1} and \cite{a-g/2}).

By varying this lagrangian with respect to the spinor field we get the corresponding field equations for the spinor as
\begin{eqnarray}
D^{2}\lambda+K^{\mu}D_{\mu}\lambda+m^{2}\lambda=0
\label{equationspinorfield}
\end{eqnarray}
which are second-order derivative field equations, with derivatives that contain contorsion and for which contorsion can be separated by writing
\begin{eqnarray}
\nabla^{2}\lambda+\frac{1}{2}\nabla_{\mu}K^{\alpha\beta\mu}\sigma_{\alpha\beta}\lambda
+K^{ij\mu}\sigma_{ij}\nabla_{\mu}\lambda+\frac{1}{4}K^{ij\mu}K_{ab\mu}\sigma_{ij}\sigma^{ab}\lambda
+m^{2}\lambda=0
\label{equationspinor}
\end{eqnarray}
identically. 

By varying the lagrangian with respect to any connection, or with respect to the contorsion we get the corresponding field equations for the spin as 
\begin{equation}
\left(K_{\mu[\alpha\beta]}+K_{[\alpha}g_{\beta]\mu}\right)=
\frac{1}{2}\left(D_{\mu}\stackrel{\neg}{\lambda}\sigma_{\alpha\beta}\lambda
-\stackrel{\neg}{\lambda}\sigma_{\alpha\beta}D_{\mu}\lambda\right)
\label{equationsspin}
\end{equation}
which relate the contorsion tensor to the spin tensor, this last tensor being written in terms of the spinor field derivatives in which contorsion can be separated apart as
\begin{eqnarray}
\nonumber
&4\left(K_{\mu\alpha\beta}-K_{\mu\beta\alpha}+K_{\alpha}g_{\beta\mu}-K_{\beta}g_{\alpha\mu}\right)
+\frac{1}{2}K^{\sigma\rho}_{\phantom{\sigma\rho}\mu}
\varepsilon_{\alpha\beta\sigma\rho}\left(i\stackrel{\neg}{\lambda}\gamma\lambda\right)-\\
&-K_{\alpha\beta\mu}\left(\stackrel{\neg}{\lambda}\lambda\right)
-2\left(\nabla_{\mu}\stackrel{\neg}{\lambda}\sigma_{\alpha\beta}\lambda
-\stackrel{\neg}{\lambda}\sigma_{\alpha\beta}\nabla_{\mu}\lambda\right)=0
\label{relationship}
\end{eqnarray}
which is a relationship that determines contorsion as an implicit function of the contorsionless derivatives of the spinor field, and which can be inverted to explicitly get contorsion in terms of the contorsionless derivatives of the spinor field itself as
\begin{eqnarray}
\nonumber
&K_{\alpha\beta\mu}
\left[(8-\stackrel{\neg}{\lambda}\lambda)^{2}
-(i\stackrel{\neg}{\lambda}\gamma\lambda)^{2}\right]
\left[(4+\stackrel{\neg}{\lambda}\lambda)^{2}
-(i\stackrel{\neg}{\lambda}\gamma\lambda)^{2}\right]=\\
\nonumber
&=\left[(8-\stackrel{\neg}{\lambda}\lambda)^{2}
-(i\stackrel{\neg}{\lambda}\gamma\lambda)^{2}\right]
(i\stackrel{\neg}{\lambda}\gamma\lambda)
(\stackrel{\neg}{\lambda}\sigma^{\sigma\rho}\nabla_{\mu}\lambda
-\nabla_{\mu}\stackrel{\neg}{\lambda}\sigma^{\sigma\rho}\lambda)
\varepsilon_{\sigma\rho\alpha\beta}+\\
\nonumber
&+2\left[(8-\stackrel{\neg}{\lambda}\lambda)^{2}
-(i\stackrel{\neg}{\lambda}\gamma\lambda)^{2}\right]
(4+\stackrel{\neg}{\lambda}\lambda)
(\stackrel{\neg}{\lambda}\sigma_{\alpha\beta}\nabla_{\mu}\lambda
-\nabla_{\mu}\stackrel{\neg}{\lambda}\sigma_{\alpha\beta}\lambda)-\\
\nonumber
&-4\left[(4+\stackrel{\neg}{\lambda}\lambda)
(8-\stackrel{\neg}{\lambda}\lambda)
-(i\stackrel{\neg}{\lambda}\gamma\lambda)^{2}\right]
(\stackrel{\neg}{\lambda}\sigma_{\sigma\theta}\nabla_{\zeta}\lambda
-\nabla_{\zeta}\stackrel{\neg}{\lambda}\sigma_{\sigma\theta}\lambda)
\varepsilon_{\alpha\beta\mu\rho}\varepsilon^{\sigma\theta\zeta\rho}+\\
\nonumber
&+8\left[(4+\stackrel{\neg}{\lambda}\lambda)
(8-\stackrel{\neg}{\lambda}\lambda)
-(i\stackrel{\neg}{\lambda}\gamma\lambda)^{2}\right]
(\stackrel{\neg}{\lambda}\sigma_{\eta\alpha}\nabla^{\eta}\lambda
-\nabla^{\eta}\stackrel{\neg}{\lambda}\sigma_{\eta\alpha}\lambda)g_{\mu\beta}-\\
\nonumber
&-8\left[(4+\stackrel{\neg}{\lambda}\lambda)
(8-\stackrel{\neg}{\lambda}\lambda)
-(i\stackrel{\neg}{\lambda}\gamma\lambda)^{2}\right]
(\stackrel{\neg}{\lambda}\sigma_{\eta\beta}\nabla^{\eta}\lambda
-\nabla^{\eta}\stackrel{\neg}{\lambda}\sigma_{\eta\beta}\lambda)g_{\mu\alpha}-\\
\nonumber
&-16(2-\stackrel{\neg}{\lambda}\lambda)
(i\stackrel{\neg}{\lambda}\gamma\lambda)
(\stackrel{\neg}{\lambda}\sigma^{\eta\rho}\nabla_{\eta}\lambda
-\nabla_{\eta}\stackrel{\neg}{\lambda}\sigma^{\eta\rho}\lambda)\varepsilon_{\alpha\beta\mu\rho}+\\
\nonumber
&+8(2-\stackrel{\neg}{\lambda}\lambda)
(i\stackrel{\neg}{\lambda}\gamma\lambda)
(\stackrel{\neg}{\lambda}\sigma^{\sigma\theta}\nabla^{\zeta}\lambda
-\nabla^{\zeta}\stackrel{\neg}{\lambda}\sigma^{\sigma\theta}\lambda)
g_{\mu\beta}\varepsilon_{\sigma\theta\zeta\alpha}-\\
&-8(2-\stackrel{\neg}{\lambda}\lambda)
(i\stackrel{\neg}{\lambda}\gamma\lambda)
(\stackrel{\neg}{\lambda}\sigma^{\sigma\theta}\nabla^{\zeta}\lambda
-\nabla^{\zeta}\stackrel{\neg}{\lambda}\sigma^{\sigma\theta}\lambda)
g_{\mu\alpha}\varepsilon_{\sigma\theta\zeta\beta}
\label{contorsion}
\end{eqnarray}
identically.

This expression can be inserted into the field equations (\ref{equationspinor}) where the contorsionless derivatives of contorsion of the field is the contorsionless derivatives of contorsionless derivatives of the field, resulting in field equations containing only contorsionless derivatives of the field itself.

\subsection{ELKOs: characteristic propagating solutions}
Because in the field equations the highest-order derivative terms are the only ones that determine the causal behaviour for the propagating wavefronts, then we consider the highest-order derivative terms alone and then we formally replace them with the normal to the propagating wavefronts $\nabla_{\mu} \rightarrow n_{\mu}$ requiring the resulting equation to be singular: thus one obtains an equation in terms of $n$ called characteristic equation, whose solutions for the vector $n$ are such that if $n$ has positive-defined norm then the fields are boosted out of the light-cone (for details about the discussion on causality see \cite{v-z}). 

Here, the characteristic equation is of the general form
\begin{eqnarray}
\mathrm{det}\left(An^{2}+C^{\mu\nu}n_{\nu}n_{\mu}\right)=0
\label{characteristicequation}
\end{eqnarray}
where
\begin{eqnarray}
\nonumber
&A=\left[(8-\stackrel{\neg}{\lambda}\lambda)^{2}
-(i\stackrel{\neg}{\lambda}\gamma\lambda)^{2}\right]
\left[(4+\stackrel{\neg}{\lambda}\lambda)^{2}
-(i\stackrel{\neg}{\lambda}\gamma\lambda)^{2}\right]\mathbb{I}+\\
\nonumber
&+\left[(8-\stackrel{\neg}{\lambda}\lambda)
(4-\stackrel{\neg}{\lambda}\lambda)(4+\stackrel{\neg}{\lambda}\lambda)
-(i\stackrel{\neg}{\lambda}\gamma\lambda)^{2}
(\stackrel{\neg}{\lambda}\lambda)\right]
\sigma_{\alpha\beta}\lambda\stackrel{\neg}{\lambda}\sigma^{\alpha\beta}+\\
&+\left[(8-\stackrel{\neg}{\lambda}\lambda)^{2}
-(i\stackrel{\neg}{\lambda}\gamma\lambda)^{2}\right]
(\frac{i\stackrel{\neg}{\lambda}\gamma\lambda}{2})
\sigma_{\alpha\beta}\lambda\stackrel{\neg}{\lambda}\sigma_{\sigma\rho}\varepsilon^{\sigma\rho\alpha\beta}
\label{matrix}\end{eqnarray}
and
\begin{eqnarray}
C^{\mu}_{\eta}=
8(2-\stackrel{\neg}{\lambda}\lambda)(i\stackrel{\neg}{\lambda}\gamma\lambda)
(\sigma_{\eta\rho}\lambda\stackrel{\neg}{\lambda}\sigma_{\sigma\theta}
+\sigma_{\sigma\theta}\lambda\stackrel{\neg}{\lambda}\sigma_{\eta\rho})\varepsilon^{\sigma\theta\rho\mu}
\label{matrixmatrices}
\end{eqnarray}
admitting for $n$ time-like solutions that would correspond to fields with acausal propagation; however, it is possible to prove that there exist conditions under which the vector $n$ is light-like, and the fields have causal propagation (\cite{f}).

\subsubsection{ELKOs: characteristic propagating solutions -\\
causality conditions}
Although knowing that there exist conditions under which the vector $n$ is light-like and hence the fields are causal is already good a result, nevertheless it would be even better to see what these conditions actually are; a long but straightforward calculation will show that the characteristic equation (\ref{characteristicequation}) with (\ref{matrix}) and (\ref{matrixmatrices}) is equivalently writable as
\begin{eqnarray}
(n^{2})^{4}\cdot\mathrm{det}E=0
\label{characteristicequationmodified}
\end{eqnarray}
where
\begin{eqnarray}
\nonumber
&E=\left[(8-\stackrel{\neg}{\lambda}\lambda)^{2}
-(i\stackrel{\neg}{\lambda}\gamma\lambda)^{2}\right]
\left[(4+\stackrel{\neg}{\lambda}\lambda)^{2}
-(i\stackrel{\neg}{\lambda}\gamma\lambda)^{2}\right]\mathbb{I}+\\
\nonumber
&+\left[(8-\stackrel{\neg}{\lambda}\lambda)
(4-\stackrel{\neg}{\lambda}\lambda)(4+\stackrel{\neg}{\lambda}\lambda)
-(i\stackrel{\neg}{\lambda}\gamma\lambda)^{2}
(\stackrel{\neg}{\lambda}\lambda)\right]
\sigma_{\alpha\beta}\lambda\stackrel{\neg}{\lambda}\sigma^{\alpha\beta}+\\
&+i\left[(8-\stackrel{\neg}{\lambda}\lambda)^{2}
-8(2-\stackrel{\neg}{\lambda}\lambda)
-(i\stackrel{\neg}{\lambda}\gamma\lambda)^{2}\right]
(i\stackrel{\neg}{\lambda}\gamma\lambda)
\gamma\sigma_{\alpha\beta}\lambda\stackrel{\neg}{\lambda}\sigma^{\alpha\beta}
\label{matrixmodified}
\end{eqnarray}
which admits for $n$ time-like solutions corresponding to fields with acausal propagation only if $\mathrm{det}E=0$ holds: however, because such condition is not verified identically, then in general the vector $n$ is light-like, the fields have causal propagation, and the result in \cite{f} has been rediscovered in the most general way.

\subsubsection{ELKOs: characteristic propagating solutions -\\
acausality over-determination}
But this result also means that whenever the necessary condition given by the vanishing of the determinant $\mathrm{det}E\equiv0$ holds we may still have acausality. Now, to see what happens when we require the vanishing of the determinant $\mathrm{det}E\equiv0$, we first notice that the matrix $E$ is always a block-diagonal matrix, and the vanishing of the determinant of a block-diagonal matrix implies the vanishing of the determinant of either one of the two block-matrices; in this case the reducibility of the representation implies that the vanishing of the determinant of both block-matrices gives the very same condition, which can be separated into real and imaginary parts in such a way that the imaginary part would eventually imply that the condition $(i\stackrel{\neg}{\lambda}\gamma\lambda)(\stackrel{\neg}{\lambda}\lambda-2)\equiv0$ holds: however in this situation the field equations would reduce to the form
\begin{eqnarray}
\left\{\begin{array}{ll}
E \cdot \nabla^{2}\lambda=H\\
\mathrm{det}E\equiv0
\end{array}\right.
\end{eqnarray}
where $H=H_{m}(\nabla\lambda)$ is a given spinor that is a function of the first-order derivatives of the field in terms of the mass as parameter, which clearly tells us that under the acausality conditions, the system of second-order differential equations is singular, and for general values of $H$ the evolution of the field is over-determined. This shows that acausality would not only be a problem for the propagation of the solutions, but also a problem for the consistency of the field equations themselves.
\section*{Conclusion}
In the present paper, we have explicitly found all the conditions of causality for ELKOs; incidentally this has allowed us to show that acausality would always let inconsistencies arise, therefore indirectly proving that causality for ELKOs is always maintained. In this way, this result provides an affirmative answer to the question of causality for ELKOs.

The fact that ELKOs are causal positively solves one of the most fundamental problems that could have possibly affected them.

Consequently, the entire ELKO theory turns out to be wholly strengthened.

\

\noindent \textbf{Acknowledgments.} I am grateful to Prof. Dharam V.~Ahluwalia and Dr. Christian G.~B\"{o}hmer for their encouragement and valuable help; I am also thankful to Prof. Giorgio Velo for enlightening discussions.

\


\end{document}